\documentclass[a4paper,fleqn,usenatbib]{mnras}
\usepackage{graphicx}
\usepackage{amssymb}
\usepackage{amsmath}
\usepackage[T1]{fontenc}
\usepackage{ae,aecompl}
\usepackage{natbib,times}
\usepackage{lscape}
\newcommand{\mc}{\multicolumn}

\begin{document}
\title[Galaxy clusters in 2MASS, WISE, and SuperCOSMOS]{A catalogue of
  clusters of galaxies identified from all sky surveys of 2MASS, WISE,
  and SuperCOSMOS}

\author[Z. L. Wen et al.]
{Z. L. Wen$^1$\thanks{E-mail: zhonglue@nao.cas.cn}, 
J. L. Han$^{1,2}$\thanks{E-mail: hjl@nao.cas.cn} 
and F. Yang$^{1,2}$
\\
$^1$National Astronomical Observatories, Chinese Academy of Sciences, 
20A Datun Road, Chaoyang District, Beijing 100012, China;\\
$^2$School of Astronomy and Space Sciences, University of the
Chinese Academy of Sciences, Beijing 100049, China\\
}

\date{Accepted XXX. Received YYY; in original form ZZZ}

\label{firstpage}
\pagerange{\pageref{firstpage}--\pageref{lastpage}}
\maketitle


\begin{abstract}
We identify 47600 clusters of galaxies from photometric data of Two
Micron All Sky Survey (2MASS), Wide-field Infrared Survey Explorer
(WISE) and SuperCOSMOS, among which 26,125 clusters are recognized for
the first time and mostly in the sky outside the Sloan Digital Sky
Survey (SDSS) area. About 90\% of massive clusters of
$M_{500}>3\times10^{14} M_{\odot}$ in the redshift range of
$0.025<z<0.3$ have been detected from such survey data, and the
detection rate drops down to 50\% for clusters with a mass of
$M_{500}\sim1\times10^{14} M_{\odot}$. Monte Carlo simulations show
that the false detection rate for the whole cluster sample is less
than 5\%. By cross-matching with ROSAT and XMM--Newton sources, we get
779 new X-ray cluster candidates which have X-ray counterparts within
a projected offset of 0.2~Mpc.
\end{abstract}


\section{Introduction}

Clusters of galaxies are the largest gravitationally bound systems in
the Universe. They are located at the knots in the cosmic web.
Galaxy clusters can be identified from observational data in optical,
X-ray and millimeter bands \citep[e.g.][]{abe58,sz72,ak83}.
In optical band, galaxy clusters have been identified based on
clustering of galaxies on the sky \citep{abe58,aco89} or clustering in
the three-dimensional space \citep{hg82,ymv+07,tes+08,
  whl09,whl12,spd+11}, or even based on galaxy colours
\citep{gy00,gsn+02,kma+07b,rrb+14}.
In X-ray images, clusters can be efficiently recognized as diffuse
extended sources. A few thousand X-ray clusters have been 
identified \citep[e.g.][]{bvh+00,bsg+04,tsl11,bcc+13,ftc+15,ltt+15,
  bcr+17}, and the largest sample of X-ray clusters in the local
universe have been found from the ROentgen SATellite (ROSAT) all sky
survey data \citep[e.g.][]{pap+11}.
Sensitive observations of cosmic microwave background (CMB) with
Wilkinson Microwave Anisotropy Probe (WMAP) and Planck satellite as
well as the ground telescopes show the distortion of CMB spectrum by
hot intracluster gas, i.e. Sunyaev-Zel'dovich (SZ) effect. Up to now,
1653 low-redshift clusters were found via the SZ effect from the
second Planck data release \citep{plancksz16} and a few hundred of
clusters up to a high redshift from the Atacama Cosmology Telescope
(ACT) and South Pole Telescope (SPT) surveys
\citep{maa+11,hhm+13,bsd+15}.

Recent years, the Sloan Digital Sky Survey \citep[SDSS, ][]{yaa+00}
provides a great data base for identification of galaxy
clusters. Galaxy clusters or groups have been identified from the SDSS
spectroscopic data \citep[e.g.,][]{mz05,bfw+06,tst+10,ttg+14} or the
five-band photometric data
\citep[e.g.,][]{gsn+02,kma+07b,whl09,hmk+10,spd+11,rrb+14,rrh+16,bsp+18}.
Currently, the largest catalogue contains 158103 clusters detected in
the sky area of 14,000 deg$^2$ \citep{whl12,wh15}. Outside of the SDSS
sky region, however, only a few thousand clusters are known
\citep[e.g.][]{abe58,aco89,dms97,bvh+00,bcc+13} and most of them have
a low redshift.

All sky survey data of the Two Micron All Sky Survey
\citep[2MASS,][]{scs+06}, Wide-field Infrared Survey Explorer
\citep[WISE,][]{wem+10} and SuperCOSMOS \citep{him01} provide a good
opportunity to identify more galaxy clusters in the sky area outside
the SDSS. In this paper, we identify 47,600 galaxy clusters by using
photometric redshifts of galaxies from the all sky data of 2MASS, WISE
and SuperCOSMOS. In Section 2, we briefly describe these survey
data. In Section 3, we describe the identification procedures for
galaxy clusters. In Section 4, we discuss the properties of this
sample of galaxy clusters, including the completeness, false detection
rate and richness estimates. The identified clusters are cross-matched
with the X-ray sources in the ROSAT and XMM--Newton surveys to find new
X-ray cluster candidates. A short summary is presented in Section 5.

Throughout this paper, we assume a flat $\Lambda$ cold dark matter
cosmology taking $H_0=70$ km~s$^{-1}$ Mpc$^{-1}$, $\Omega_m=0.3$ and
$\Omega_{\Lambda}=0.7$.

\section{Galaxy data for cluster identification}

Three data sets of all sky surveys are used for cluster identification. 
The 2MASS\footnote{http://www.ipac.caltech.edu/2mass/} is an all-sky
survey in three near-infrared bands \citep{scs+06}: $J$ (1.25 $\mu$m),
$H$ (1.65 $\mu$m) and $K_s$ (2.17 $\mu$m) with the 10$\sigma$
magnitude limits of 15.8, 15.1, and 14.3 mag for point sources,
respectively. More than 100 million sources are detected, and the
Extended Source Catalogue (XSC) contains about one million galaxies up
to a limit of $J=15.0$ mag \citep{jcc+00}.
The WISE\footnote{http://irsa.ipac.caltech.edu/Missions/wise.html}
observes the whole sky in four mid-infrared bands \citep{wem+10}: $W1$
(3.4 $\mu$m), $W2$ (4.6 $\mu$m), $W3$ (12 $\mu$m) and $W4$ (22 $\mu$m)
with 5$\sigma$ magnitude limits of 17.1, 15.7, 11.5, and
7.7 mag in the Vega system for point sources, respectively.
The SuperCOSMOS\footnote{http://www-wfau.roe.ac.uk/sss/} data are
measured from the photographic plates of the Palomar Observatory Sky
Survey-II (POSS-II) in the north and the United Kingdom Schmidt
Telescope (UKST) in the south, which include optical photometric
magnitudes in three bands ($B$, $R$ and $I$) with limits of $B\sim21$
mag and $R\sim19.5$ mag \citep{phb+16}.
Obviously, the WISE survey in the $W1$ and $W2$ goes much deeper than
the 2MASS but is comparable with the SDSS for galaxies \citep{ydt+13}.

By combining the data of 2MASS, WISE and SuperCOSMOS, \citet{bjp+14}
estimated photometric redshifts for about one million 2MASS XSC
galaxies (2MPZ\footnote{http://surveys.roe.ac.uk/ssa/TWOMPZ}) by using
the artificial neural network (ANNz) approach, which have an
uncertainty of $\sigma_z=0.015$ and a median redshift of $z\sim0.1$.
\citet{bpj+16} further estimated the photometric redshifts of about 20
million galaxies\footnote{http://ssa.roe.ac.uk/WISE$\times$SCOS}
covering 28,000 square degree of the sky in the WISE and SuperCOSMOS
surveys by using the ANNz approach again. They removed stars and
quasars by the colour cuts of $W1-W2$ and $R-W2$, and found that the
so-obtained `galaxies' with $B<21$, $R<19.5$ and $13.8<W1<17$ at
high Galactic latitudes is approximately 95\% pure (i.e. mixed with
stars up to 5\% of the total number of listed `galaxies') and 90\%
complete (i.e. about 10\% of real galaxies may be missing due to the
colour cuts). The photometric redshifts of the galaxies have a median
of of 0.2 and an uncertainty of $\sigma_z=0.033(1+z)$.

\begin{figure}
\centering
\includegraphics[width = 72mm]{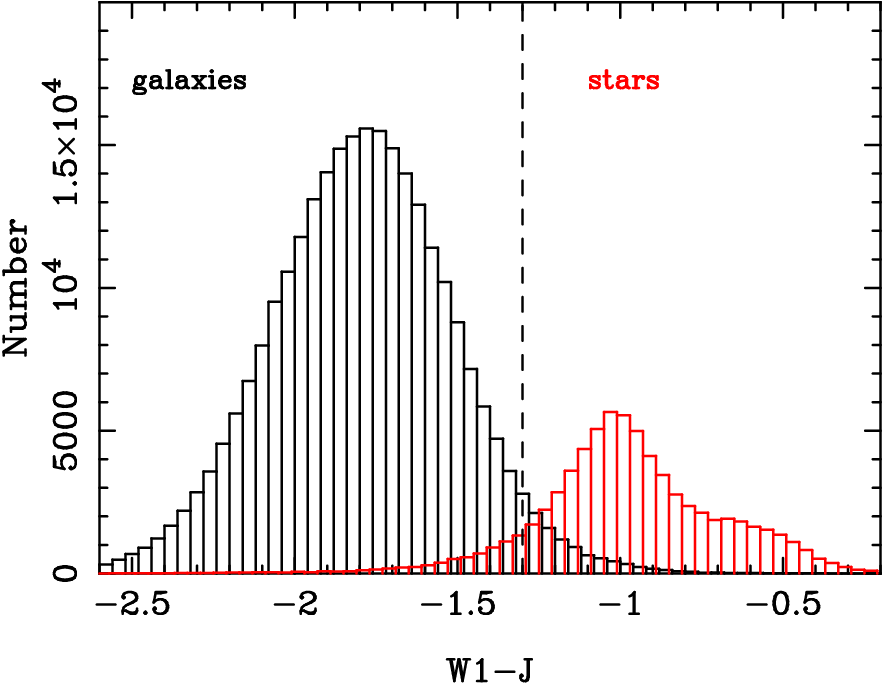}
\caption{Stars (red histogram) and galaxies (black histogram) are
  distributed with two peaks along the colour of $W1-J$. The data here
  are WISE and 2MASS objects in a SDSS sky area of 1500 square degree
  at a middle Galactic latitude. Clearly the criterion of $W1-J>-1.3$
  can be taken to remove most of contaminating stars.}
\label{starremov}
\end{figure}

\begin{figure}
\centering
\includegraphics[width = 72mm]{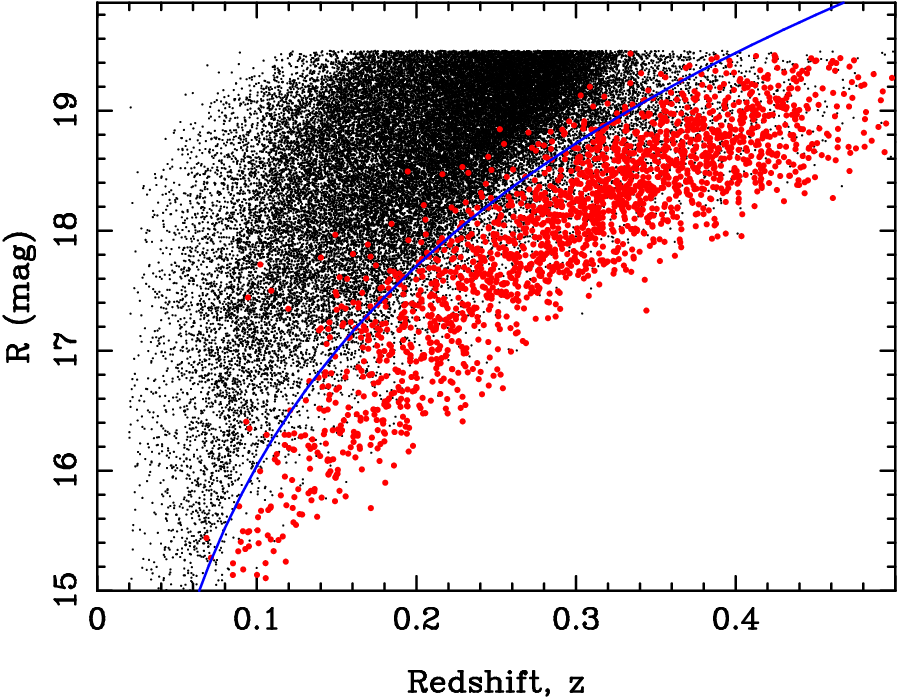}
\includegraphics[width = 72mm]{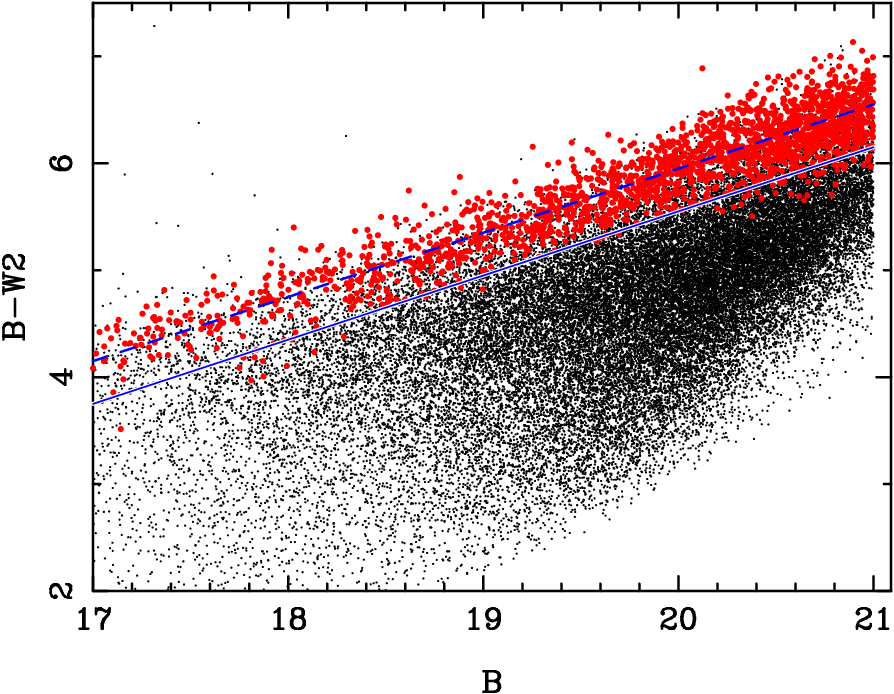}
\caption{Bright red galaxies can be taken as BCG candidates, with
  criteria shown in the magnitude--redshift ($R$--$z$) distribution
  (upper panel) and the colour--magnitude ($B-W2$ versus $B$)
  distribution (lower panel). The big (red) dots are 2000 known
  BCGs of randomly selected clusters from \citet{wh15} in a sky region
  near the Galactic pole region, and small dots are 40,000 galaxies
  randomly selected in the region. The blue solid lines are the
  magnitude cut (upper panel) given by equation~(\ref{bcgmz})
  and colour cut (lower panel) given by equation~(\ref{bcgcm})
  to select BCG candidates. The dashed line in the lower panel is the
  linear fit to the distribution of the BCGs.}
\label{BCGcandi}
\end{figure}

\begin{figure*}
\centering
\includegraphics[width = 150mm]{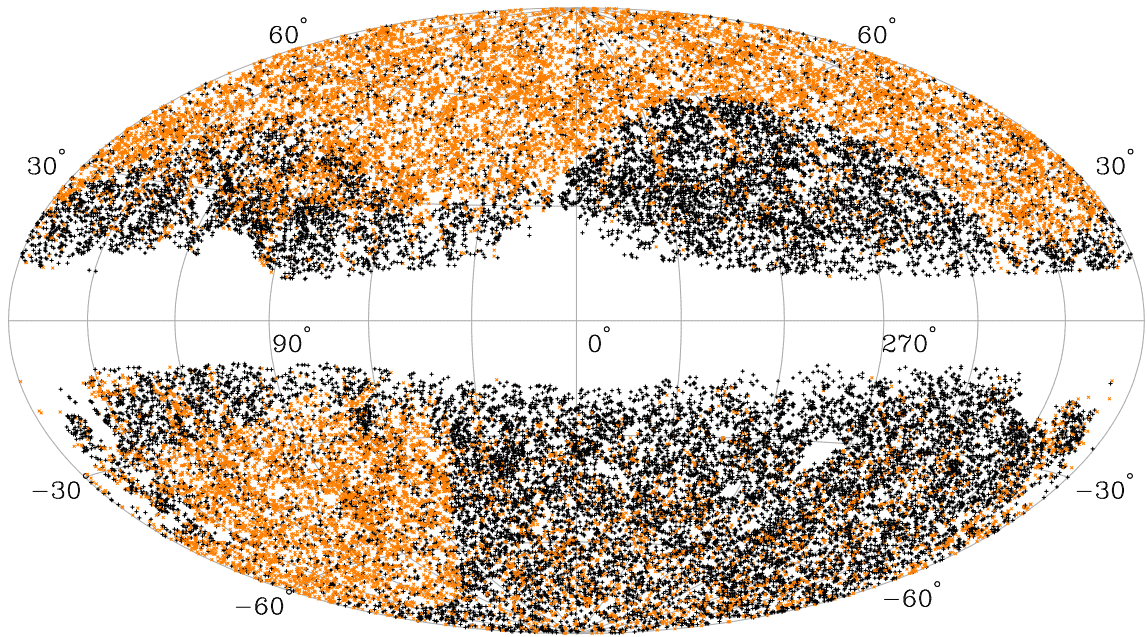}
\caption{The sky distribution of 47,600 identified galaxy clusters in
  the Galactic coordinates. The orange `$\times$' stands for 21,475
  previously known clusters and black `$+$' for 26,125 newly
  identified clusters.}
\label{clustersky}
\end{figure*}

The main data set for identification of galaxy clusters is the sample
of 20 million WISE$\times$SuperCOSMOS galaxies (see Section 3) which
have photometric redshifts estimated by \citet{bpj+16}. To make the
galaxy sample as complete as possible at the bright end, 0.8 million
bright 2MPZ galaxies of $W1<13.8$ in the same 28,000 square degree sky
coverage \citep{bjp+14} are further supplementally included. For
common galaxies appearing in both catalogues, the redshifts from
2MPZ catalogue are adopted.
We also notice that the `galaxies' in the catalogue of
\citet{bpj+16} are still contaminated with some bright stars, which
can be distinguished by using the colour of $W1-J$
\citep{xwh14,ks15}. As shown in Fig.~\ref{starremov}, stars and
galaxies are distributed with two peaks, which can be roughly
separated by a criterion of $W1-J=-1.3$. Here the recognized stars and
galaxies are taken from the SDSS data with a limit of $r=21.5$
\citep{lgi+01}. This colour cut would cause 1.4\% of total galaxies
omitted, but the contaminating stars can be reduced to 3\% at the
Galactic latitudes of $|b|\ge 30^\circ$ and 6.5\% at the Galactic
latitudes of $|b|<30^\circ$. With this process, about 1.9 million
sources are removed from the WISE$\times$SuperCOSMOS catalogue of
\citet{bpj+16}. Finally, 18.4 million galaxies from 2MASS, WISE and
SuperCOSMOS are used for identification of galaxy clusters in the
following.

\section{Galaxy clusters identified from SuperCOSMOS, WISE and 2MASS}

Photometric redshift can be used to discriminate member galaxies of
clusters and field galaxies during identification of galaxy clusters
\citep[e.g.][]{wyz+06,whl09,spd+11,whl12}. A galaxy cluster usually
contains one or more giant luminous member galaxies, and one of them
is the brightest cluster galaxy (BCG). We first select a sample of BCG
candidates, and then identify clusters as being overdensity regions of
galaxies around them.

\subsection{Find out BCG candidates}
\label{findbcg}

We take the following steps to select red luminous galaxies as being
BCG candidates.
First of all, BCG candidates must be bright enough to be detected in
the 2MASS $J$ band and the WISE $W1$ band, so that the contamination
of stars can be minimized by the colour cut of $W1-J<-1.3$. Secondly,
they must be bright in the $R$ band of SuperCOSMOS. In general, BCGs
have a narrow distribution of absolute magnitude or colour
\citep{san72,pl95}.  The plot for $R$-band magnitude against redshift
for galaxies and BCGs of known SDSS clusters \citep[from ][]{wh15}
clearly shows in the upper panel of Fig.~\ref{BCGcandi} that about
95\% of BCGs have an evolution-corrected absolute magnitude of
\begin{equation}
M_R\equiv R-25-5\log\,D_L(z)-[e(z)+k(z)] < -22.3, 
\end{equation}
or the apparent magnitude of
\begin{equation}
R<-22.3+25+5\log\,D_L(z)+[e(z)+k(z)].
\label{bcgmz}
\end{equation}
Here $D_L(z)$ is luminosity distance of a galaxy, and $e(z)$ and
$k(z)$ are the terms for evolution-correction and $K$-correction which
can be given as $e(z)+k(z)=0.24 z$ at $z<0.4$ according to the
stellar population synthesis model of \citet{bc03}.
Thirdly, the very red colour of BCGs can be quantified in the 
colour--magnitude diagram of $B-W2$ versus $R$ by using the data of
SuperCOSMOS and WISE. As shown in the lower panel of
Fig.~\ref{BCGcandi}, about 95\% of the BCGs have the colour cut of
\begin{equation}
B-W2>0.6B-6.45.
\label{bcgcm}
\end{equation}
With the criteria of equations~(\ref{bcgmz}) and (\ref{bcgcm}), we
obtain 0.76 million galaxies as BCG candidates. Their spectroscopic
redshifts are taken from the 2MPZ catalogue \citep{bjp+14} if
available, otherwise photometric redshifts are used.

\begin{figure}
\centering
\includegraphics[width = 71mm]{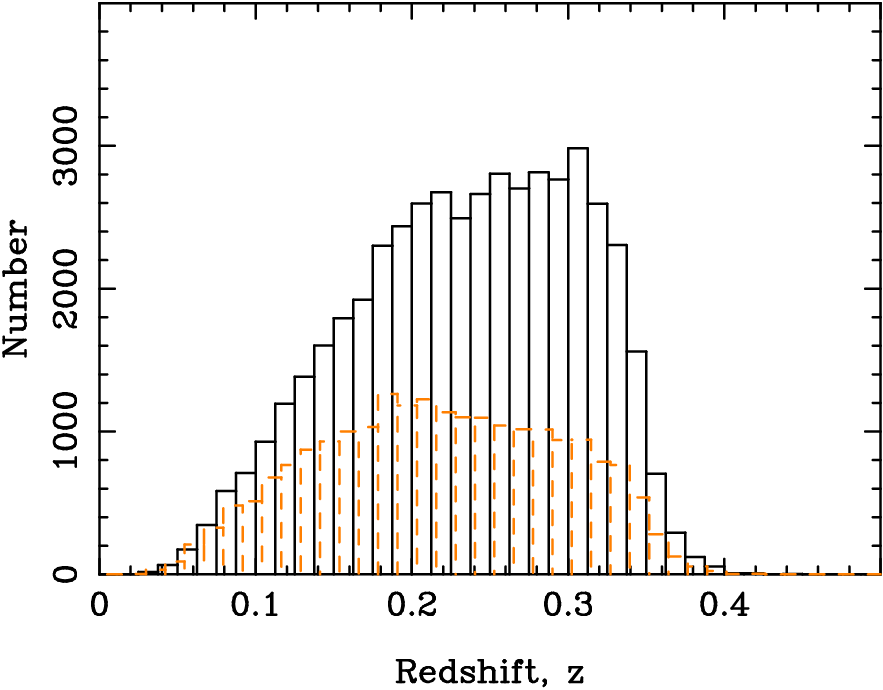}
\caption{The redshift distribution of 47,600 identified clusters of
  galaxies, with 21,475 previously known clusters indicated by the
  dashed line.}
\label{hist_zc}
\end{figure}

\begin{figure}
\centering
\includegraphics[width = 71mm]{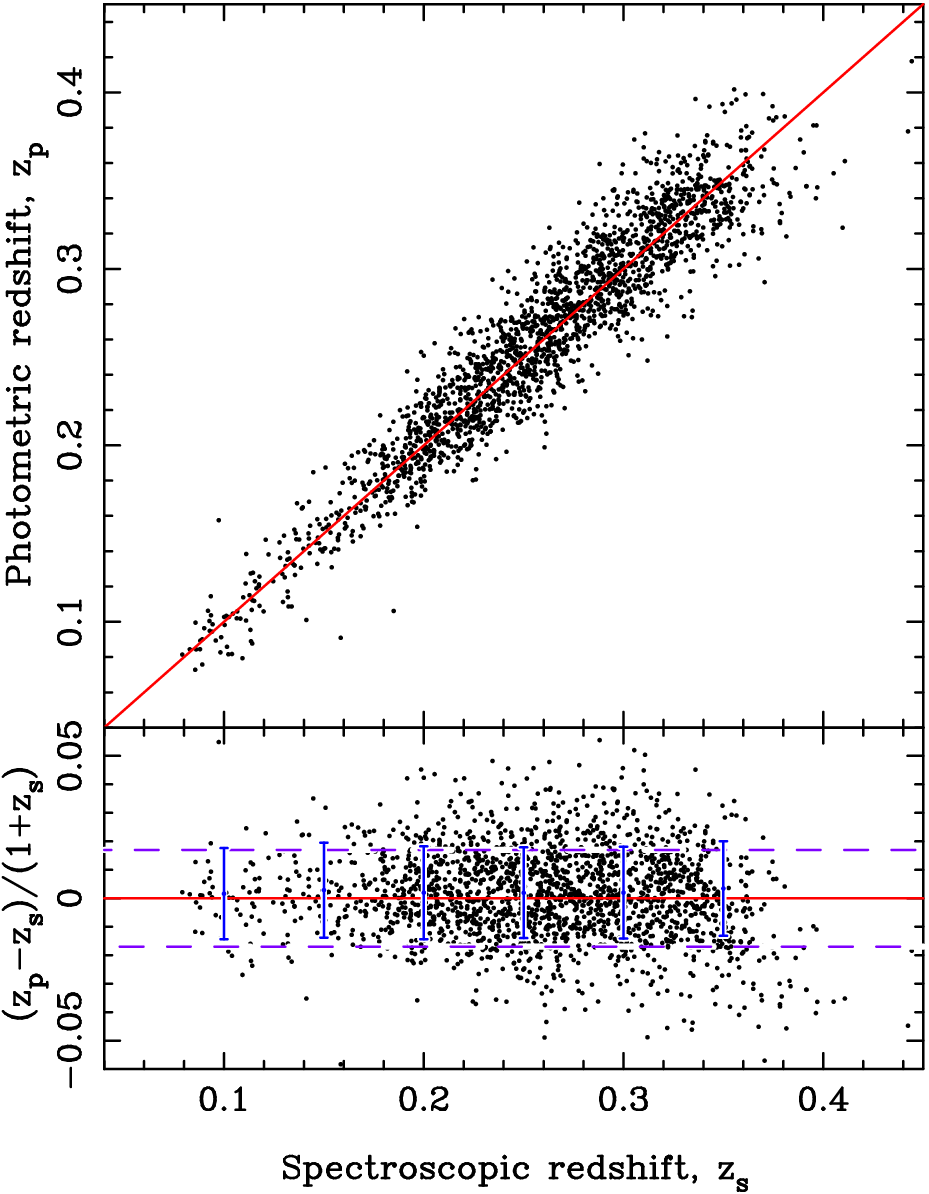}
\caption{Comparison between BCG photometric redshifts ($z_p$) with
  their spectroscopic redshifts ($z_s$) obtained by SDSS \citep{dr12}
  for 6030 identified clusters. The accuracy of cluster photometric
  redshifts is $\sigma=0.017(1+z)$ as indicated by the dashed lines in
  the lower panel.}
\label{photoz}
\end{figure}

\subsection{Find out galaxy clusters}
\label{findcluster}

Galaxy clusters stand out as the overdensity of galaxies around the
BCG candidates. Referring our previous papers \citep{whl09,wh11}, we
identify galaxy clusters with the following steps:

1. Get the number count of galaxies around each of the BCG candidates we
have selected above. Since the spectroscopic or photometric redshift,
$z$, of a BCG candidate is known, we can get the count of galaxies,
$N_{\rm 0.5 Mpc}$, within a projected radius of 0.5~Mpc from the BCG
candidate in the redshift slice of $z\pm0.05(1+z)$. Here the radius of
0.5~Mpc is about 2 times of the core size of a rich cluster
\citep{gbg+95,amk+98}, and the thickness of redshift slice is 1.5
times of the scatter of the uncertainty of estimated galaxy photometric
redshift \citep{bpj+16}, so that about 85\% of member galaxies of a
possible cluster should be included.

2. Estimate the local `background' and `fluctuation' of the number
count of galaxies. For each cluster candidate, the average of the
local background, $\langle N_{\rm 0.5 Mpc}\rangle$, should be
subtracted. Here $\langle N_{\rm 0.5 Mpc}\rangle$ is estimated locally
within the same redshift slice of $z\pm0.05(1+z)$ but in an annulus of
projected distance between 2--4 Mpc from the BCG candidate. The
fluctuation of the number count, $\sigma_{N_{\rm 0.5 Mpc}}$, is also
estimated as being the deviation of the number count of galaxies at
1000 random positions in the redshift slice.

3. Calculate the `signal-to-noise ratio' to find an overdensity
region. The signal-to-noise ratio is defined as being ${\rm
  S/N}=(N_{\rm 0.5 Mpc}-\langle N_{\rm 0.5 Mpc}\rangle)/\sigma_{N_{\rm
    0.5 Mpc}}$. A larger signal-to-noise ratio means a higher overdensity
of galaxies, which is the direct indication of a true cluster. To keep
a low false detection rate ($<$5\%, see discussions in
Section~\ref{comp_puri}), we set the threshold as being ${\rm
  S/N}\ge4$.

4. Clean the overdensity entries to make a list of galaxy clusters. It
is possible that two or more luminous member galaxies in a rich
cluster can be recognized as the BCG candidates, so that one cluster
can be sort out twice or more times through the above procedures. We
therefore perform the friends-of-friends algorithm \citep{hg82} to
merge them into one cluster if they have a redshift difference smaller
than $\sim 2\,\sigma_z=0.075(1+z)$ and a projected distance
smaller than 1~Mpc (a typical radius of a cluster). The BCG with the
highest ${\rm S/N}$ is then adopted for such a combined cluster.

\begin{table*}
\begin{minipage}{160mm}
\caption[]{21,475 known clusters of galaxies recognized from the
  survey data of SuperCOSMOS, WISE and 2MASS.}
\begin{center}
\begin{tabular}{rrrrccrrl}
\hline
\mc{1}{c}{Name}&\mc{1}{c}{RA} & \mc{1}{c}{Declination} & \mc{1}{c}{$z$} & \mc{1}{c}{flag$_z$} & 
\mc{1}{c}{$R_{\rm BCG}$} & \mc{1}{c}{${\rm S/N}$} & \mc{1}{c}{$R_{L*}$} & \mc{1}{c}{Other catalogues}\\
\mc{1}{c}{(1)} & \mc{1}{c}{(2)} & \mc{1}{c}{(3)} & \mc{1}{c}{(4)} & \mc{1}{c}{(5)} & 
\mc{1}{c}{(6)} & \mc{1}{c}{(7)} & \mc{1}{c}{(8)} & \mc{1}{c}{(9)} \\
\hline
 J000000.6$+$321233 &   0.00236 & $ 32.20922$ & 0.0913 &  0 & 15.21 &  7.75 & 53.94 & Abell,WHL\\
 J000002.3$+$051718 &   0.00946 & $  5.28823$ & 0.1747 &  0 & 16.45 &  6.08 & 39.37 & WHL      \\
 J000003.6$+$314708 &   0.01490 & $ 31.78561$ & 0.0933 &  0 & 15.52 &  7.71 & 24.27 & WHL      \\
 J000006.6$+$315235 &   0.02761 & $ 31.87626$ & 0.2125 &  0 & 17.15 &  4.77 & 32.33 & WHL      \\
 J000006.7$-$212400 &   0.02792 & $-21.39997$ & 0.1686 &  0 & 16.07 &  5.36 & 35.79 & Abell    \\
 J000007.6$+$155003 &   0.03179 & $ 15.83417$ & 0.1528 &  1 & 16.13 & 11.29 & 39.51 & Abell,WHL\\
 J000008.0$+$343316 &   0.03353 & $ 34.55436$ & 0.2863 &  0 & 18.04 &  4.76 & 30.28 & WHL  \\
 J000009.4$+$211655 &   0.03903 & $ 21.28196$ & 0.3026 &  0 & 17.92 &  8.67 & 35.42 & WHL  \\
 J000012.6$+$103806 &   0.05232 & $ 10.63496$ & 0.1794 &  0 & 16.73 &  6.56 & 51.38 & WHL  \\
 J000014.0$+$063329 &   0.05825 & $  6.55797$ & 0.2197 &  0 & 17.59 &  4.51 & 26.45 & WHL  \\
\hline
\end{tabular}
\end{center}
{Note. 
Column 1: Cluster name with J2000 coordinates of cluster. 
Column 2: RA (J2000) of cluster BCG (degree).
Column 3: Declination (J2000) of cluster BCG (degree). 
Column 4: cluster redshift, with a flag in Column (5):
`0' for photometric redshift and `1' for spectroscopic redshift. 
Column 6: $R$-band magnitude of BCG.
Column 7: ${\rm S/N}$ of the overdensity for cluster recognition.
Column 8: cluster richness.
Column 9: other catalogues containing the cluster: 
Abell \citep{abe58,aco89}, Zwicky \citep{zwi61}, BM78 \citep{bm78}, GHO \citep{gho86}, EDCC \citep{lnc+92}, APM \citep{dms97}, CE \citep{gsn+02}, NSCS \citep{ldg+04}, C4 \citep{mnr+05}, TES+06 \citep{tes+06}, maxBCG \citep{kma+07b}, YMV+07 \citep{ymv+07}, NSC \citep{gld+09}, GMBCG \citep{hmk+10}, 
AMF \citep{spd+11,bsp+18}, MCXC \citep{pap+11}, GMB+11, \citep{gmb11}, WH11 \citep{wh11}, WHL \citep{whl12,wh15}, MSPM \citep{shh+12}, TTL12 \citep{ttl12}, RXSC \citep{cbn13}, redMaPPer \citep{rrb+14,rrh+16}, CAMIRA \citep{ogu14}, 
SPT \citep{bsd+15}, SWXCS \citep{ltt+15}, DAB+15, \citep{dab+15}, LCS \citep{bsb+15}, PSZ2 \citep{plancksz16}, TKT16 \citep{tkt+16}, XLSSC \citep{pcg+16},
and KDR2 \citep{rpb+17}.\\
(This table is available in its entirety in a machine-readable form.)
}
\label{tab1}
\end{minipage}
\end{table*}

\begin{table*}
\begin{minipage}{160mm}
\caption[]{26,125 newly identified clusters of galaxies from the
  survey data of SuperCOSMOS, WISE and 2MASS.}
\begin{center}
\begin{tabular}{rrrrccrr}
\hline
\mc{1}{c}{Name}&\mc{1}{c}{RA} & \mc{1}{c}{Declination} & \mc{1}{c}{$z$} & \mc{1}{c}{flag$_z$} & 
\mc{1}{c}{$R_{\rm BCG}$} & \mc{1}{c}{${\rm S/N}$} & \mc{1}{c}{$R_{L*}$}\\
\mc{1}{c}{(1)} & \mc{1}{c}{(2)} & \mc{1}{c}{(3)} & \mc{1}{c}{(4)} & \mc{1}{c}{(5)} & 
\mc{1}{c}{(6)} & \mc{1}{c}{(7)} & \mc{1}{c}{(8)} \\
\hline
WHY J000001.3$-$561854 &   0.00538 & $-56.31503$ & 0.3212 &  0 & 18.43 &  6.85 & 15.74 \\                     
WHY J000004.3$-$483442 &   0.01780 & $-48.57825$ & 0.3555 &  0 & 18.08 &  9.17 & 36.05 \\                     
WHY J000006.1$-$075319 &   0.02528 & $ -7.88872$ & 0.1153 &  0 & 16.21 &  4.80 & 12.27 \\                     
WHY J000007.6$+$420725 &   0.03181 & $ 42.12375$ & 0.1891 &  0 & 17.32 &  4.26 & 23.34 \\                     
WHY J000008.6$-$150422 &   0.03564 & $-15.07269$ & 0.2571 &  0 & 18.04 &  4.79 & 21.69 \\                     
WHY J000010.0$-$140805 &   0.04183 & $-14.13463$ & 0.2574 &  0 & 17.59 &  4.11 & 31.40 \\                    
WHY J000010.4$-$305008 &   0.04338 & $-30.83558$ & 0.1707 &  0 & 16.62 &  4.18 & 10.52 \\                     
WHY J000012.8$-$120153 &   0.05353 & $-12.03148$ & 0.2406 &  1 & 16.74 &  5.91 & 26.41 \\                     
WHY J000017.6$-$290246 &   0.07347 & $-29.04622$ & 0.2564 &  0 & 17.88 &  5.87 & 18.93 \\                     
WHY J000018.2$+$392145 &   0.07589 & $ 39.36263$ & 0.1248 &  0 & 16.04 &  4.42 & 14.02 \\
\hline
\end{tabular}
\end{center}
{Note.
Column 1: Cluster name with J2000 coordinates of cluster. 
Column 2: RA (J2000) of cluster BCG (degree).
Column 3: Declination (J2000) of cluster BCG (degree). 
Column 4: cluster redshift, with a flag in Column (5):
`0' for photometric redshift and `1' for spectroscopic redshift. 
Column 6: $R$-band magnitude of BCG.
Column 7: ${\rm S/N}$ of the overdensity for cluster recognition.
Column 8: cluster richness.\\
(This table is available in its entirety in a machine-readable form.)
}
\label{tab2}
\end{minipage}
\end{table*}

We finally identify 47,600 clusters through the above procedures.
Table~\ref{tab1} lists 21,475 previously known clusters (see
Section~\ref{match_cata}) and Table~\ref{tab2} lists 26,125 newly
identified clusters with a prefix of WHY to indicate their nature of
new discovery in this paper. The sky distribution of these galaxy
clusters is shown in Fig.~\ref{clustersky}, which follows the
distribution of WISE$\times$SuperCOSMOS galaxies given by
\citet{bpj+16}. The redshifts of galaxy clusters are mostly within
$0.05<z<0.36$ as shown in Fig.~\ref{hist_zc}. By comparing them with
the SDSS spectroscopic data \citep{dr12} available for the BCGs of
6030 clusters we identified (see Fig.~\ref{photoz}), we find that the
accuracy of cluster photometric redshifts is about $0.017(1+z)$.

\begin{figure}
\centering
\includegraphics[width = 75mm]{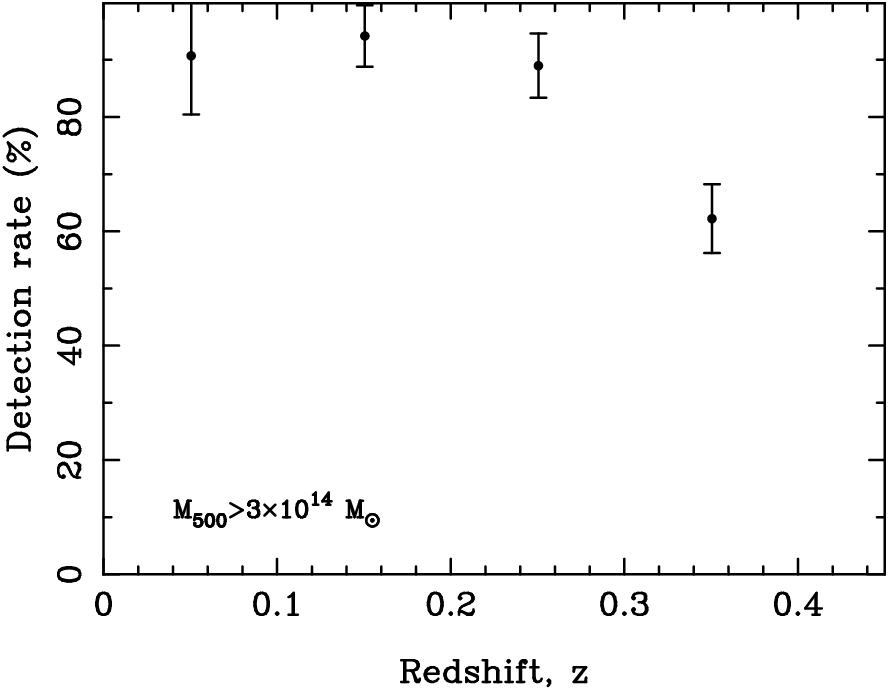}
\includegraphics[width = 75mm]{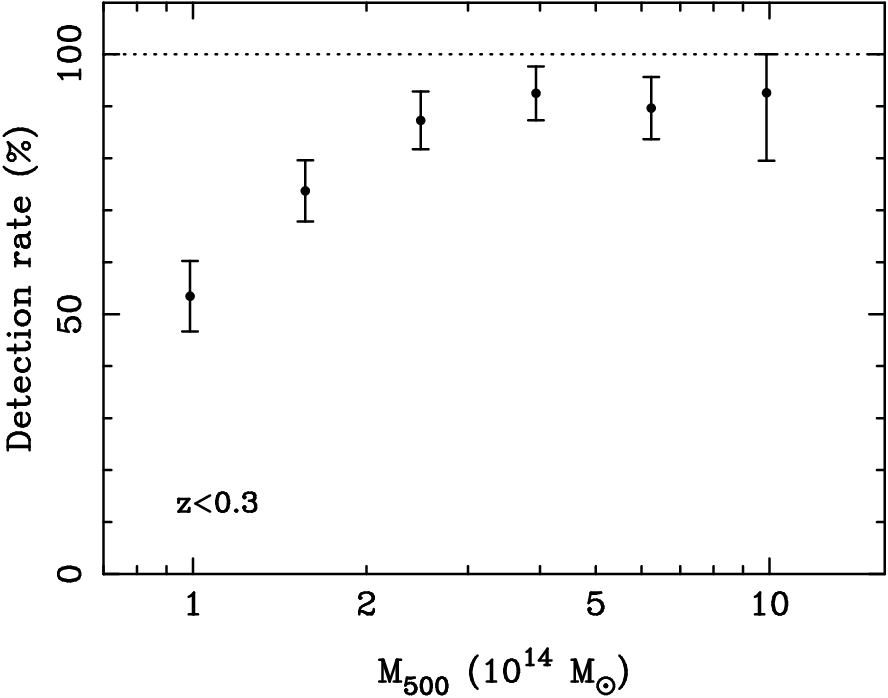}
\caption{Detection rate of 1613 known massive clusters as a function
  of redshift (upper panel) and cluster mass (lower panel).}
\label{matchrate}
\end{figure}

\section{Discussions}

\subsection{Detection rate}
\label{comp_puri}

Massive clusters with many bright member galaxies can easily be
detected through the above cluster identification procedures. To show
the dependence of cluster detection rate on redshift and cluster mass,
we collect a sample of 1613 massive clusters within the sky area in
Fig.~\ref{clustersky}, which have previously been detected from the
ROSAT X-ray survey \citep{vbe+09,mae+10,pap+11}, the XMM--Newton survey
\citep{tsl11} and the Planck SZ survey \citep{plancksz16}. These X-ray
and SZ samples are complete either on X-ray flux density or the SZ CMB
detection threshold, so that they are independent of optical/infrared
cluster identifications. The fraction of these X-ray and SZ clusters
detected in our cluster list can be regarded as an indicator for
completeness of the identified clusters, as shown by \citet{hmk+10},
\citet{rr14} and \citet{ogu14}, because there is no significant
difference on galaxy distributions in the X-ray detected and
non-detected clusters \citep{pbb+07b}.

Among the 1613 clusters, we get 1235 (77\%) matches within a redshift
difference of $0.05(1+z)$ and a separation of $1.5\,r_{500}$, here
$r_{500}$ is the previously known radius of massive clusters in
literature within which the mean density of a cluster is 500 times of
the critical density of the universe. Fig.~\ref{matchrate} shows the
detection rate of these X-ray and SZ clusters as a function of
redshift $z$ and cluster mass $M_{500}$, i.e. the cluster mass within
$r_{500}$ rescaled to those of \citet{vbe+09} as done by
\citet{wh15}. In the redshift range of $z<0.3$, about 90\% massive
clusters of $M_{500}>3\times10^{14}~M_{\odot}$ can be detected, and
the detection rate drops quickly down to 50\% for clusters with a mass
of $M_{500}\sim1\times10^{14}~M_{\odot}$. Therefore, clusters with a
lower mass obviously have a much lower detection rate.

In the above cluster identification procedure, the inherent assumption
is that clusters are centred on the potential BCGs. It is possible that
the BCGs are not located at cluster centres \citep{svy+11,hll+15,
  oll+17}. The miscentring can induce the number of discriminated
member galaxies and hence the `${\rm S/N}$'
underestimated. Rich clusters may have enough member galaxies around
the BCGs and still can be identified above the threshold of ${\rm
  S/N}\ge4$. Nevertheless, poor clusters may not. To investigate the
effect of miscentring on completeness, we carry out a simulation to
shift the locations of the BCGs by a projected length of 0.2 Mpc (see
Section~\ref{xray}) in a random direction and then try to search
galaxy clusters with the above procedure with the same ${\rm S/N}$ 
threshold. We find that about 22\% (10,423 of 47600) of clusters
cannot be identified due to a low ${\rm S/N}$ of ${\rm S/N}<4$.

\begin{figure}
\centering
\includegraphics[width = 70mm]{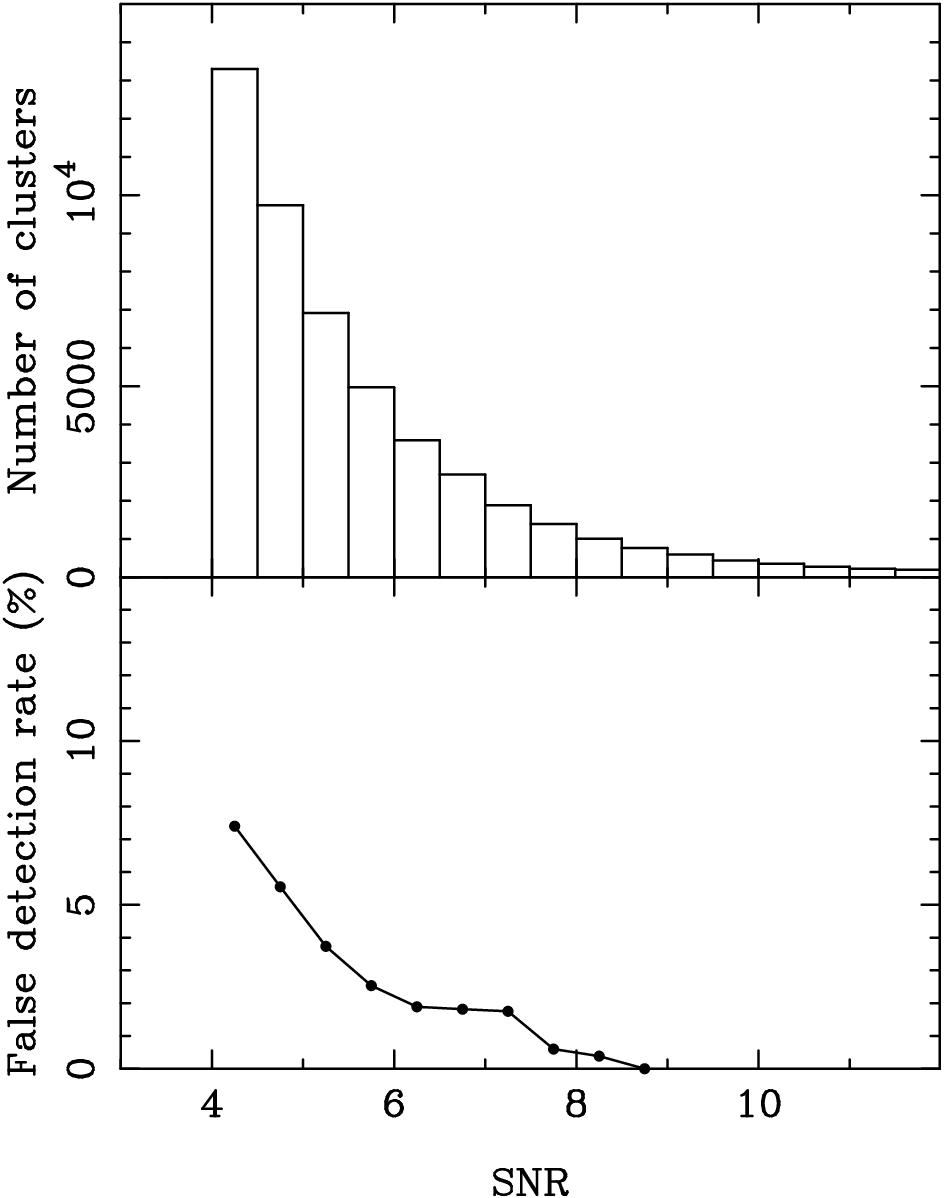}
\caption{The number distribution of identified galaxy clusters and the
false detection rate as a function of ${\rm S/N}$.}
\label{false-rate}
\end{figure}

On the other hand, the projection effect of large scale structure may
induce possible false detections of clusters. We estimate the false
detection rate by Monte Carlo simulations following many previous
authors \citep[e.g.,][]{gsn+02,kma+07b,hmk+10,whl12}. First, we
discard all recognized member galaxy candidates of 47,600 identified
clusters from the catalogue of 18.4 million galaxies. Then, ten mock
samples of rest galaxies are generated by randomly shuffling the
photometric redshifts and magnitudes of all galaxies, so that the
two-dimensional distribution of galaxies are kept as the same as the
real sample, but the detected clusters have been eliminated. By using
these mock samples, we search galaxy clusters with the steps discussed
Section~\ref{findcluster} with the same threshold of ${\rm
  S/N}\ge4$. Any detected `clusters' from such mock samples
therefore can be regarded as false detections due to projection
effect. For each mock sample, a false detection rate is calculated as
being the ratio between the number of false detected clusters from the
mock data and the number of detected clusters in the real data. To
minimize the random noise, we get an average from ten mock samples.
The false detection rate is 4.5\% for the whole cluster sample, but
varies with the `${\rm S/N}$'. It increases from nearly 0
at ${\rm S/N}>8.5$ to about 7.5\% in the bin of $4.0<{\rm S/N}<4.5$ to
(see Fig.~\ref{false-rate}).

\begin{figure}
  \centering
\includegraphics[width = 70 mm]{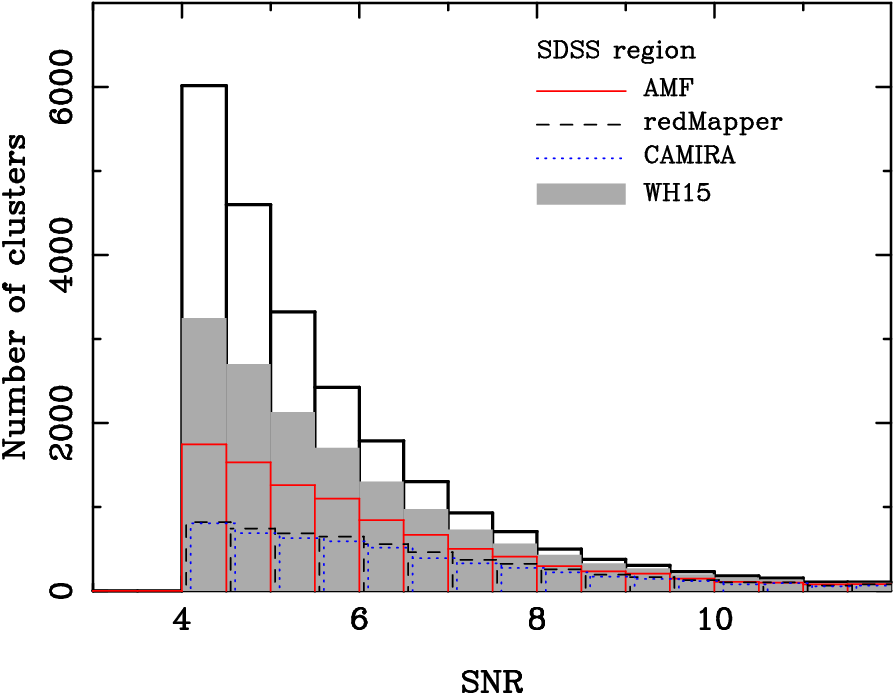}
\includegraphics[width = 70 mm]{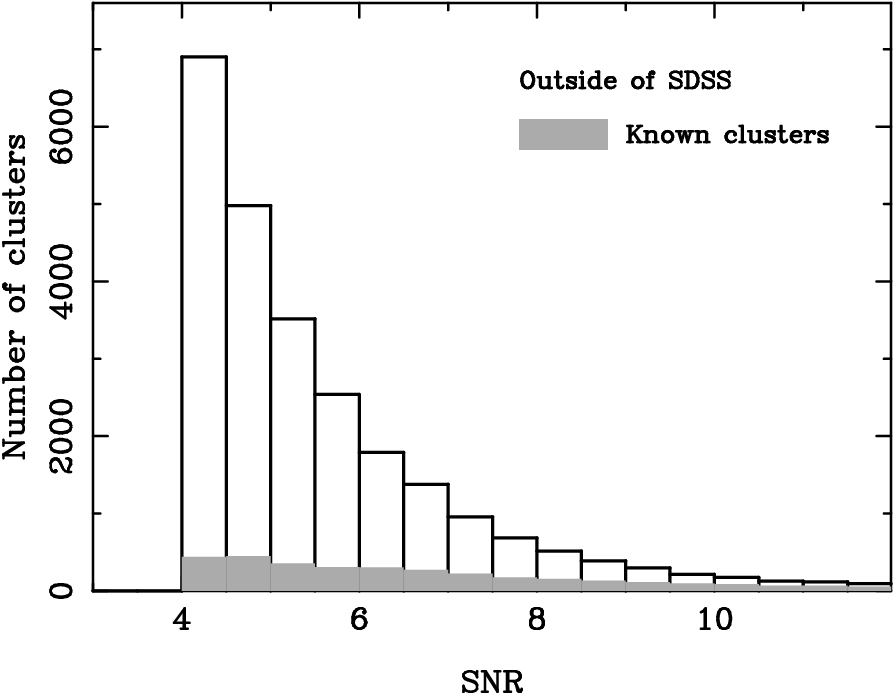}
\caption{Most galaxy clusters in the SDSS sky area have 
  been identified previously from the SDSS data (the upper panel), but outside
  the SDSS area, most of galaxy clusters are newly identified in this
  work.}
\label{match_pre}
\end{figure}
\subsection{Clusters in previous cluster catalogues}
\label{match_cata}

We cross-match the identified galaxy clusters with clusters in
previous cluster catalogues to find out how many clusters have been
previously known. We work in two regions, the sky region of SDSS
coverage and the sky region outside the SDSS.

In the sky region of SDSS coverage, we have got 23,366 clusters
identified from the 2MASS, WISE and SuperCOSMOS survey data, but a
large number of galaxy clusters there have been identified previously
from the SDSS photometric data, for example, 13,823 clusters in the
maxBCG catalogue \citep{kma+07b}, 39,668 clusters in the WHL09
catalogue \citep{whl09}, 55,424 clusters in the GMBCG catalogue
\citep{hmk+10}, 69,173 clusters in the AMF catalogue \citep{spd+11}
and 46,479 clusters in the updated AMF catalogue \citep{bsp+18},
25,325 clusters in the redMaPPer catalogue \citep{rrb+14} and 26,311
clusters in the updated redMaPPer catalogue \citep{rrh+16}, and 71,743
clusters in the CAMIRA catalogue \citep{ogu14}.
Cross-matching with the largest catalogue containing 158,103 clusters
\citep[WHL,][]{whl12,wh15} shows that 66\% of 23,366 clusters can be
matched with the WHL clusters within a redshift difference
$<0.05(1+z)$ and a projected separation $<$1 Mpc (see
Fig.~\ref{match_pre}). By adding not-overlapped clusters in other
catalogues such as the GMBCG, AMF, redMaPPer and CAMIRA, we find that
in total 18,138 galaxy clusters (78\% of 23,366) are previously known,
which means that only 5228 clusters ($\sim22\%$) in the SDSS region
are newly identified.

Outside the SDSS region, we have 24,234 galaxy clusters identified
from the 2MASS, WISE and SuperCOSMOS survey data. Previously only a
small number of galaxy clusters have been identified from either a
shallow optical \citep[e.g.,][]{aco89} or infrared surveys
\citep{kwh+03,bk12}, or the full sky X-ray survey 
\citep[e.g.,][]{pap+11}, or the CMB SZ effect data \citep{plancksz16}.
After taking out 3337 known clusters in this sky region from previous
catalogues \citep{zwi61,bm78,aco89,lnc+92,dms97,tes+06,pap+11,cbn13,
  ltt+15,bsd+15,bsb+15,tkt+16,plancksz16,pcg+16}, we conclude that
20,897 clusters are recognized for the first time (see
Fig.~\ref{match_pre}).

Therefore, in addition to the 21,475 known clusters, we get in total
26,125 galaxy clusters newly identified from the 2MASS, WISE and
SuperCOSMOS survey data (see Fig.~\ref{clustersky}).

\begin{figure}
\centering
\includegraphics[width = 65mm]{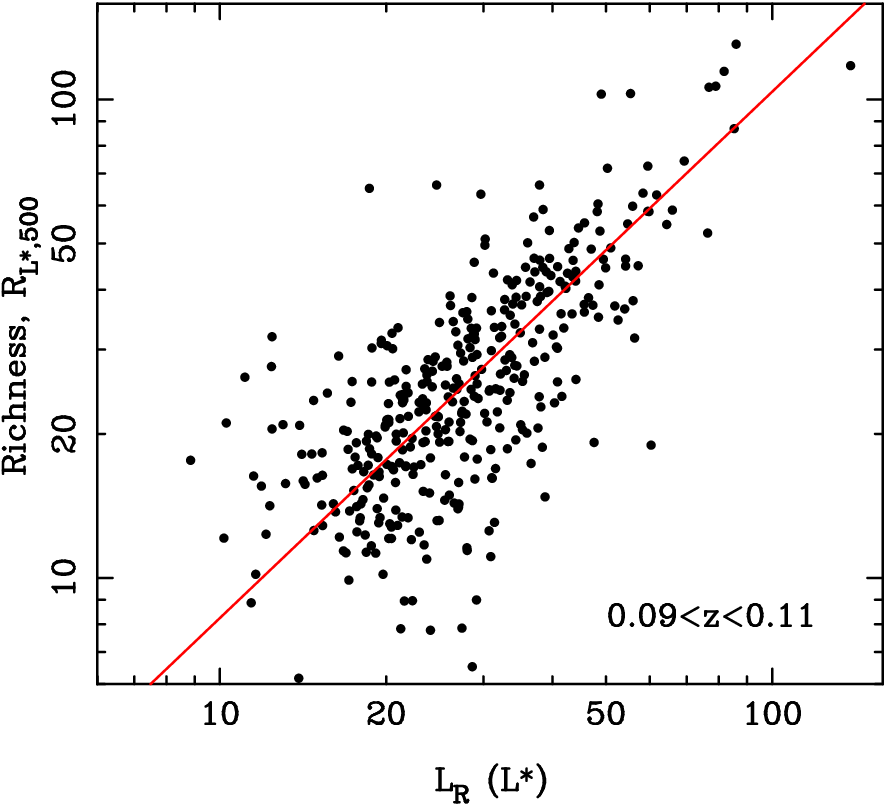}
\includegraphics[width = 65mm]{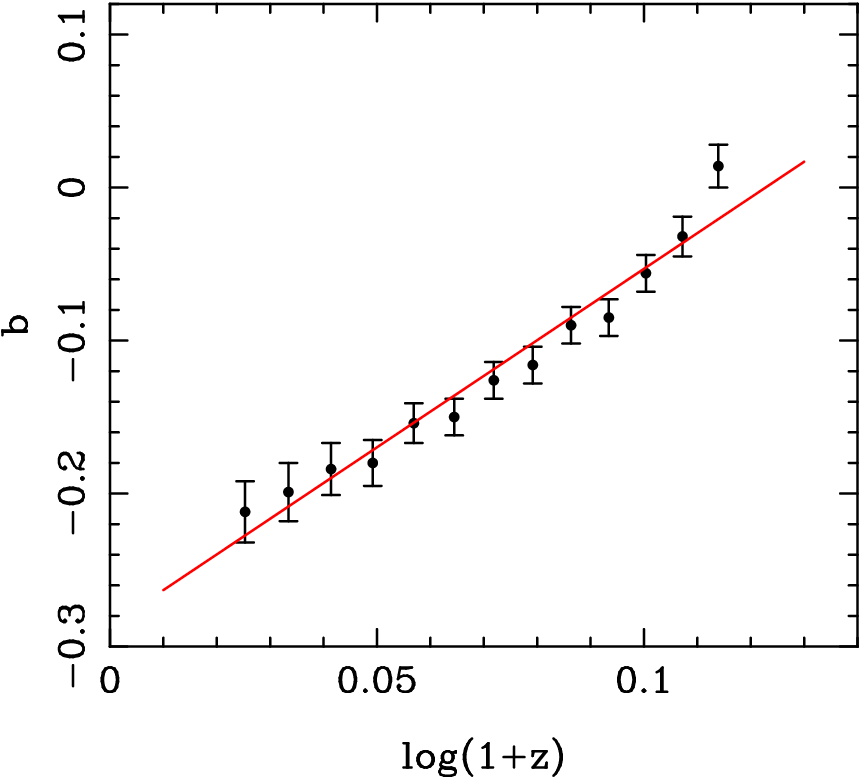}
\caption{The $R$-band total luminosity of member galaxies within a
  radius of 1~Mpc from BCGs in a small redshift ranges is closely
  related to the cluster richness of \citet{wh15} as shown
  in upper panel and described by equation~(\ref{rich_b}). The
  intersection $b$ for the scaling relation is redshift-dependent, as
  shown in the lower panel.}
\label{rich_wh15}
\end{figure}

\subsection{Richness estimation}

Richness is one of basic properties of galaxy clusters, which
describes how many member galaxies or how much mass a cluster
possesses. Richness is therefore often taken as a mass proxy. For
optically identified clusters, richness can be simply defined as the
total number of member galaxies brighter than a luminosity threshold
or magnitude limit \citep{kma+07b,hmk+10,rrb+14} or even the total
luminosity of recognized member galaxies \citep{spd+11,whl12,wh15}.

Identified from the flux-limited galaxy sample of WISE and
SuperCOSMOS, galaxy clusters at lower redshifts should contain more
member galaxies than those at higher redshifts. Therefore, the total
numbers of member galaxies or their total luminosities are obviously
biased to smaller values at high redshift, so that they cannot
directly be taken as a true cluster richness. Here we try to make
corrections to the total luminosities according to previously
calibrated richness of identified clusters \citep{xwh14}.

\begin{figure}
\centering
\includegraphics[width = 70mm]{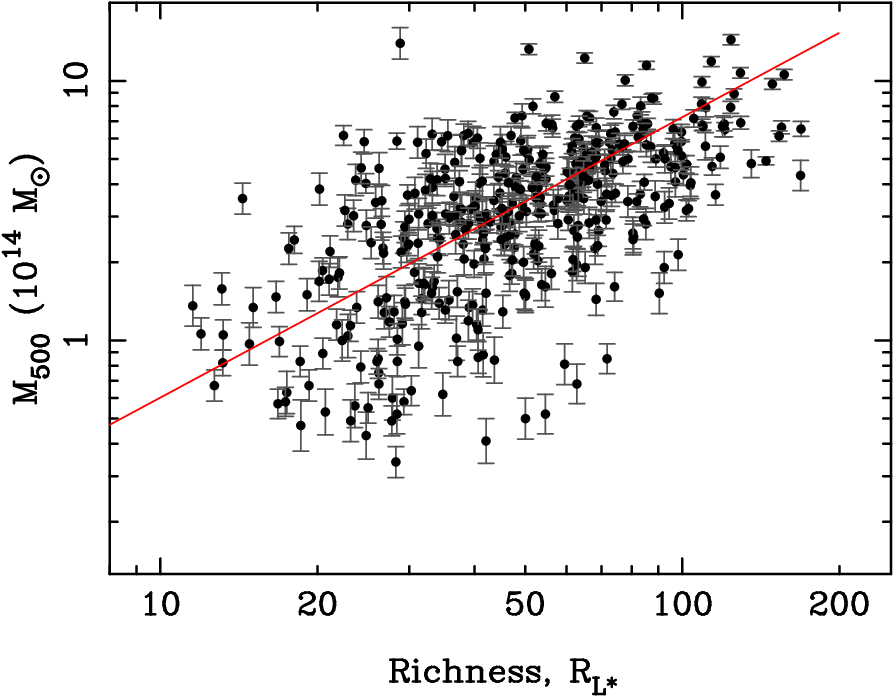}
\caption{Correlation between the derived richnesses according to
  equation~(\ref{richness}) and cluster masses for 413 galaxy clusters
  compiled in \citet{wh15}.}
\label{rich_mass}
\end{figure}

For each cluster, we first calculate the total $R$-band luminosities,
$L_R$, of member galaxies within 1 Mpc and a photometric redshift
slice of $z\pm0.05(1+z)$ with a proper background subtraction
\citep{whl12}. The $R$-band data of galaxies are taken from the
SuperCOSMOS survey. We then define the cluster richness, $R_{\rm L*}$
in unit of galaxy characteristic luminosity $L*$, as being
\begin{equation}
R_{L*}\propto L_R^{\alpha}\,(1+z)^{\beta},
\end{equation}
where $\alpha$ and $\beta$ are the power indices for luminosities and
the corrections of redshift dependence, respectively. To derive the
values of $\alpha$ and $\beta$, we get a sample of matched clusters in
Table~\ref{tab1} with those in \citet{wh15} within a projected
separation of 0.5~Mpc and a redshift difference of 0.05. Noticed that
the richness $R_{L*,500}$ given in \citet{wh15} has been calibrated
with known cluster masses estimated by X-ray or SZ measurements and
has no dependence on redshift. For the matched clusters in each of
many small redshift ranges, the correlation (such as one in the upper
panel of Fig.~\ref{rich_wh15}) can be best fitted by a power law
\begin{equation}
  \log (R_{L*,500})=a\log(L_R)+b(z),
  \label{rich_b}
\end{equation}
with a slope of $a=1.10\pm0.03$ and an intersection of $b(z)$ that is
redshift-dependent: $ b(z) =(2.32\pm0.28)\log(1+z)-(0.28\pm0.03)$ (see
the lower panel of Fig.~\ref{rich_wh15}). The richness $R_{L*,500}$ is
therefore related to $L_R$ by
\begin{equation}
R_{L*,500}=10^{-0.28\pm0.03}L_R^{1.1\pm0.03}(1+z)^{2.32\pm0.28}.
\end{equation}
Therefore, we simply define the richness of identified clusters in
this paper as
\begin{equation}
R_{L*}=0.52L_R^{1.1}(1+z)^{2.32}.
\label{richness}
\end{equation}
The richness values for all clusters are calculated and listed in Column
(8) in Table~\ref{tab1} and Table~\ref{tab2},
which are reasonably correlated with cluster masses from X-ray and SZ
measurements compiled in \citet{wh15}, as shown in
Fig.~\ref{rich_mass}. Adopting a slope of 1.08 for the richness--mass
relation \citep{wh15}, we find that the deviation of logarithmic mass
from the best fitting is about 0.25, which is comparable to those for
the maxBCG \citep{kma+07b}, GMBCG \citep{hmk+10} and AMF richnesses
\citep{spd+11}.

\begin{table*}
\begin{minipage}{160mm}
\caption[]{779 new X-ray cluster candidates and 711 known X-ray
  clusters from cross-matching with ROAST and XMM--Newton sources.}
\begin{center}
\begin{tabular}{rrrcccccl}
\hline
\mc{1}{c}{Cluster name} & \mc{1}{c}{RA} & \mc{1}{c}{Declination} & \mc{1}{c}{$z$} & \mc{1}{c}{Name of X-ray source}
& \mc{1}{c}{$f_X$} & \mc{1}{c}{$\delta_{f_X}$} & \mc{1}{c}{$r_p$} & \mc{1}{c}{Other catalogues}  \\
\mc{1}{c}{(1)} & \mc{1}{c}{(2)} & \mc{1}{c}{(3)} & \mc{1}{c}{(4)} & \mc{1}{c}{(5)} & 
\mc{1}{c}{(6)} & \mc{1}{c}{(7)} & \mc{1}{c}{(8)} & \mc{1}{c}{(9)} \\
\hline
 WHY J000108.9$-$123309 & 0.28721 & $-12.55259$ & 0.2451 & 2RXS J000109.9$-$123240 & 0.410 & 0.150 & 0.123 &      \\      
     J000141.4$-$154045 & 0.42262 & $-15.67913$ & 0.1308 & 2RXS J000140.1$-$154041 & 0.840 & 0.200 & 0.044 & MCXC \\      
     J000158.5$+$120358 & 0.49367 & $ 12.06612$ & 0.1930 & 2RXS J000155.7$+$120354 & 0.950 & 0.180 & 0.128 & MCXC \\      
     J000311.6$-$060531 & 0.79843 & $ -6.09183$ & 0.2332 & 2RXS J000311.1$-$060442 & 1.700 & 0.260 & 0.180 & MCXC \\      
     J000349.6$+$020359 & 0.95685 & $  2.06652$ & 0.0954 & 2RXS J000350.3$+$020344 & 2.200 & 0.280 & 0.031 & MCXC \\      
     J000524.0$+$161309 & 1.34998 & $ 16.21920$ & 0.1120 & 2RXS J000521.9$+$161324 & 1.380 & 0.180 & 0.068 & MCXC \\
                        &         &             &        &                         &       &       &       &      \\
     J000242.7$-$343935 & 0.67802 & $-34.65976$ & 0.1160 & 3XMM J000243.0$-$343943 & 0.160 & 0.021 & 0.020 &           \\
     J000309.4$-$295140 & 0.78896 & $-29.86099$ & 0.0720 & 3XMM J000300.1$-$295149 & 0.005 & 0.001 & 0.164 &           \\
 WHY J000647.3$-$344254 & 1.69701 & $-34.71505$ & 0.1917 & 3XMM J000646.6$-$344258 & 0.046 & 0.023 & 0.028 &           \\
 WHY J001418.8$-$302125 & 3.57844 & $-30.35692$ & 0.2317 & 3XMM J001419.5$-$302136 & 0.017 & 0.007 & 0.055 &           \\
     J002745.8$+$261626 & 6.94081 & $ 26.27401$ & 0.3662 & 3XMM J002745.1$+$261616 & 0.392 & 0.049 & 0.071 & MCXC      \\
     J004335.1$+$010111 &10.89621 & $  1.01961$ & 0.1957 & 3XMM J004334.1$+$010107 & 0.004 & 0.003 & 0.049 & 2XMMi-SDSS\\
\hline
\end{tabular}
\end{center}
{Note. 
Column (1): Cluster name;  
Column (2): RA (J2000) of cluster (degree); 
Column (3): Declination (J2000) of cluster (degree); 
Column (4): cluster redshift; 
Column (5): Name of X-ray source with J2000 coordinates; 
Column (6): X-ray flux ($10^{-12}$ erg\,cm$^{-2}$\,s$^{-1}$) in the 0.1--2.4 keV 
band for the ROSAT sources and in the 2.0--4.5 keV for the XMM--Newton sources;
Column (7): uncertainty of the X-ray flux ($10^{-12}$ erg\,cm$^{-2}$\,s$^{-1}$);
Column (8): projected offset between the X-ray source and the cluster (Mpc);
and Column (9): previous X-ray cluster catalogue: MCXC \citep{pap+11}, 2XMMi-SDSS \citep{tsl11,tsl13,tsl14},
WYS+14 \citep{wys+14}, WHL09 \citep{whl09}, SWXCS \citep{ltt+15}, and XLSSC \citep{pcg+16}.
The entries without the prefix of WHY in the column (1) are known clusters in previous optical, SZ or X-ray 
cluster catalogues (see Table~\ref{tab1}).\\
(This table is available in its entirety in a machine-readable form.)
}
\label{tab3}
\end{minipage}
\end{table*}

\begin{figure}
\centering
\includegraphics[width = 68mm]{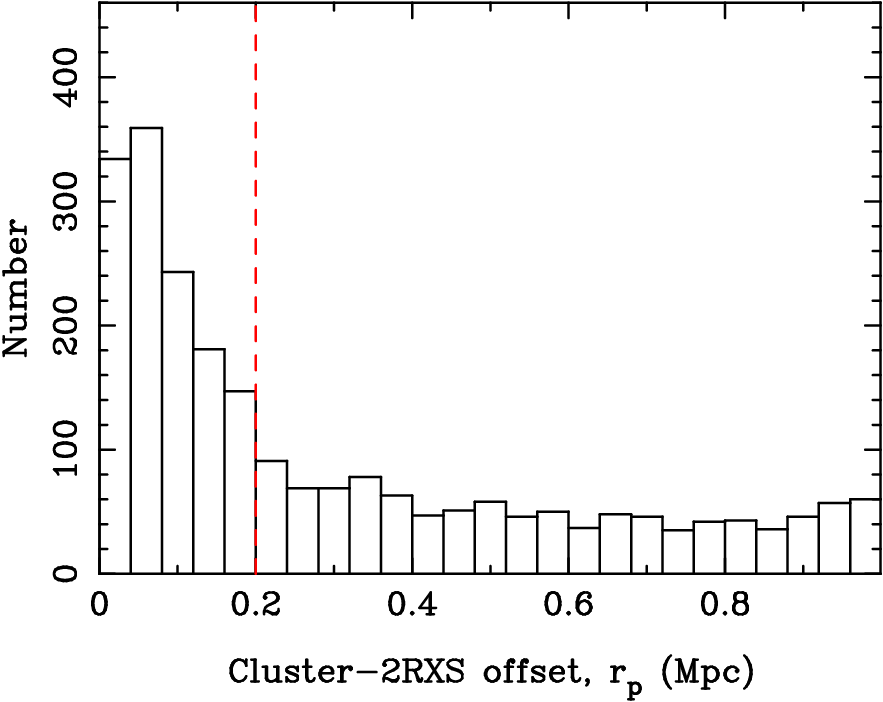}
\includegraphics[width = 68mm]{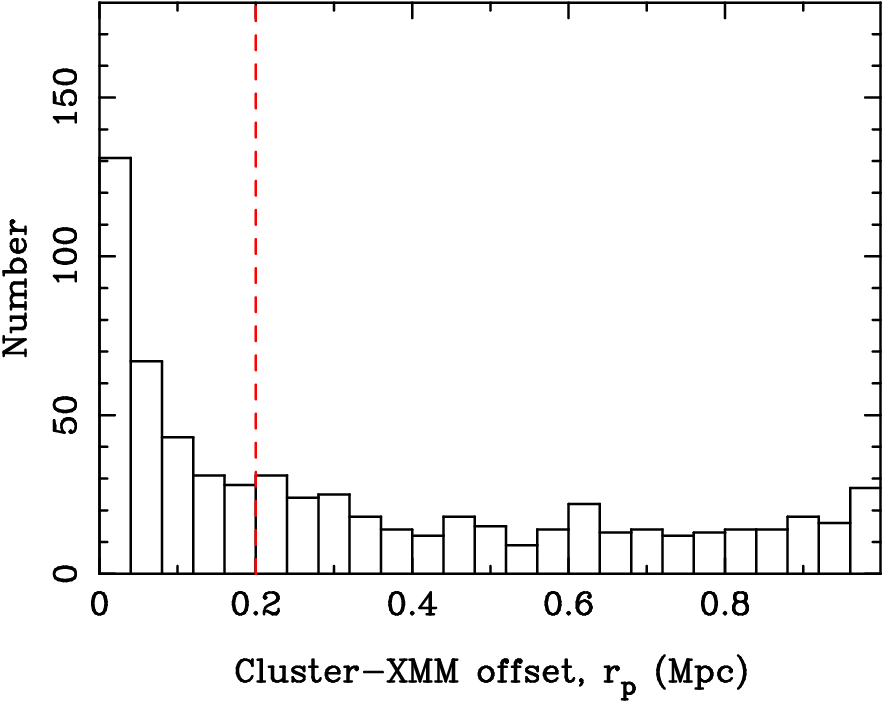}
\caption{Distribution of projected offset between the identified
  clusters and X-ray sources from the ROSAT catalogue (upper
    panel) and the XMM--Newton catalogue (lower panel).
  The excess at $r_p<0.2$ Mpc indicates the association of
  X-ray sources with clusters.}
\label{x_cluster}
\end{figure}

\subsection{X-ray emission of galaxy clusters}
\label{xray}

Galaxy clusters possess hot intracluster medium and can be detected as
extended X-ray sources. Here we check if the identified clusters from
the 2MASS, WISE and SuperCOSMOS survey data are X-ray sources by
cross-matching them with ROSAT and XMM-Newton X-ray source catalogues.

The ROSAT All Sky Survey (RASS) performed imaging observations of the
whole sky in the 0.1--2.4 keV energy band \citep{tru82}. The latest
second ROSAT source catalogue (2RXS) contains 135,118 X-ray detections
above a likelihood threshold of 6.5 and a flux limit of $10^{-13}$
erg\,cm$^{-2}$\,s$^{-1}$ \citep{bft+16}.
More than one thousand X-ray clusters have been identified previously 
as extended sources in the RASS image data: the Northern ROSAT All Sky
(NORAS) cluster catalogue contains 378 clusters \citep{bvh+00}; The
ROSAT--ESO Flux Limited X-ray (REFLEX) cluster catalogue contains 447
clusters with a flux limit of $3\times 10^{-12}$
erg\,cm$^{-2}$\,s$^{-1}$ in the Southern sky \citep{bsg+04}. Recently,
the REFLEX and NORAS cluster samples have been updated, which contain
915 and 860 clusters, respectively \citep[][data not public
  yet]{bcc+13,bcr+17}.
We cross-match galaxy clusters in Table~\ref{tab1} and
Table~\ref{tab2} with the all sky ROSAT X-ray sources in
\citet{bft+16} to find new X-ray cluster candidates. As shown in
Fig.~\ref{x_cluster}, the distribution of ROSAT X-ray sources around
the clusters shows an obvious number excess at the projected offset of
$r_p<0.2$~Mpc, which suggests the association of these X-ray sources
with galaxy clusters. If X-ray sources are randomly distributed
without any association, the source count should increase with
$r_p^2$. From 35,241 X-ray sources with a detection likelihood greater
than 10 and with a hardness-ratio in the ranges of $0<{\rm HR1}<1.0$
and $-0.2<{\rm HR2}<0.8$ \citep[see][]{vab+99}, we find that there are
ROSAT X-ray sources within a projected offset of $r_p<0.2$ Mpc for
1267 clusters. Among them, 600 are known X-ray clusters
\citep{pap+11,wys+14,ltt+15} or cluster candidates \citep{whl09}, and
the rest 667 clusters are therefore new X-ray cluster candidates as
listed in Table~\ref{tab3}.

Currently the XMM--Newton is one of the most sensitive X-ray
observatories in the energy range of 0.2--12 keV \citep{jla+01} with a
large field of view of 30 arcmin \citep{taa+01}. Individual pointed
observations provide serendipitous X-ray surveys of the sky, and about
50--100 X-ray sources can be detected from every single pointing
\citep{wsf+09} with a typical position error of 2 arcsec.
We noticed that the latest XMM--Newton X-ray source catalogue,
3XMMI--DR6, lists 486,440 unique X-ray sources down to a flux limit of
$10^{-14}$ erg\,cm$^{-2}$\,s$^{-1}$ from 9160 XMM-Newton observations
\citep{rww+16}. Among them, 37746 extended X-ray sources are above a
detection likelihood greater than 10 and have a size larger than 6
arcsec.
We also cross-match our galaxy clusters with these X-ray sources. We
first remove all (either extended or point) X-ray sources within 5
arcsec of WISE objects, which are probably bright stars or active
galactic nuclei, and then find out the extended XMM-Newton sources
around galaxy clusters listed in Table~\ref{tab1} and
Table~\ref{tab2}. Extended XMM--Newton X-ray sources are found within
$r_p<0.2$ Mpc from 376 clusters, as shown in Fig.~\ref{x_cluster}.
Among them, 153 sources have already been included in the 2RXS, and
another 111 sources come from known X-ray clusters
\citep{pap+11,tsl11,tsl13,tsl14,wys+14,ltt+15,pcg+16} or X-ray cluster
candidates \citep{whl09}. The rest 112 extended XMM--Newton X-ray
sources are therefore new X-ray cluster candidates as listed in
Table~\ref{tab3}.

Combining the ROSAT and XMM--Newton sources together, we get
779 new X-ray cluster candidates in total.

\begin{figure}
\centering
\includegraphics[width = 70mm]{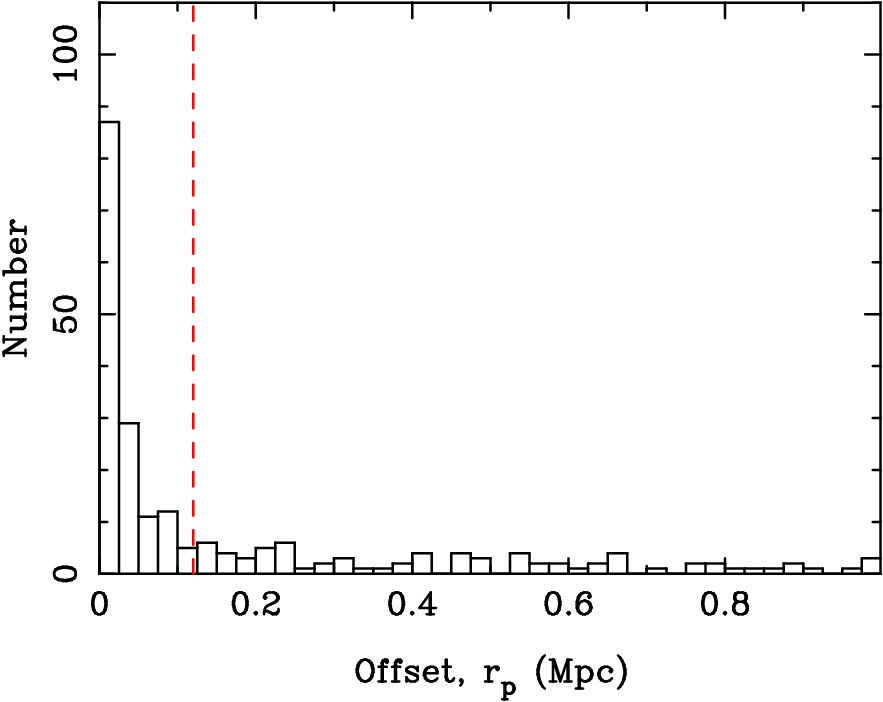}
\caption{Distribution of projected offset between the BCG locations
  and X-ray centres for 219 clusters. The BCGs in 144 clusters with 
  a small offset less than 0.12 Mpc can be regarded as true central
  galaxies in clusters.}
\label{offset}
\end{figure}

Recent studies showed that about 20--40\% of BCGs are not central
galaxies \citep{svy+11,hll+15,oll+17}. Here we also check if the BCGs of
our identified clusters are true central galaxies based on the
offsets between BCG locations and the centres in high resolution X-ray
images of clusters \citep[see][]{ogu14,rr14,rrh+16}. We use three
lists of high resolution X-ray data: 239 massive clusters in the
ACCEPT catalogue observed by Chandra \citep{cdv+09}, 94 clusters with
high-resolution Chandra follow-up observations among the 238 massive
clusters in the Mantz catalogue selected from ROSAT survey
\citep{mae+10}, and 503 clusters from the XMM--Newton Cluster Survey
\citep{mrh+12}. Merging three catalogues gives 478 unique clusters at
redshifts $z<0.4$, of which 219 clusters are matched by our cluster
catalogue.
Fig.~\ref{offset} shows the distribution of projected offset between
the BCG locations and X-ray centres. It peaks at the offset of
$r_p\sim0$ and has a standard deviation of $\sigma_{\rm p}=0.04$~Mpc,
suggesting that most BCGs are well centred. We find that 66\% (144
of 219) of the BCGs in our catalogue are central galaxies within the
offset of 3\,$\sigma_{\rm p}=0.12$~Mpc. The offset distribution
becomes flat at $r_p>0.2$ Mpc, suggesting that these BCGs probably are
miscentred. 

\section{Summary}

By using the photometric redshift data for about 18.4 million galaxies
of 2MASS, WISE and SuperCOSMOS obtained by \citet{bjp+14} and
\citet{bpj+16}, we identify 47,600 galaxy clusters in the sky survey
area of 28,000 square degree. The BCG candidates are first selected by
using the criteria of galaxy magnitude and colour. Clusters are then
identified as the overdensity regions of galaxy counts around the BCG
candidates. The detection rate is as high as 90\% for massive clusters
of $M_{500}>3\times10^{14} M_{\odot}$ in the redshift range of
$0.025<z<0.3$, and drops down to 50\% for clusters with a mass of
$M_{500}\sim 1\times 10^{14} M_{\odot}$. Among 47,600 galaxy clusters, 26,125
clusters are identified for the first time which are mostly located in
the sky outside the SDSS area. Monte Carlo
simulations show that the false detection rate is less than 5\% in
general. By cross-matching with ROSAT and XMM--Newton sources, we find
779 new X-ray cluster candidates.

\section*{Acknowledgements}

We thank the referee for valuable comments that helped to improve the
paper.
The authors are supported by the National Natural Science Foundation
of China (Grant No. 11473034), the Key Research Program of the Chinese
Academy of Sciences (Grant No. QYZDJ-SSW-SLH021) and the Young
Researcher Grant of National Astronomical Observatories, Chinese
Academy of Sciences.
This work makes use the data from the Two Micron All Sky Survey,
Wide-field Infrared Survey Explorer, SuperCOSMOS and SDSS.

\bibliographystyle{mnras}
\bibliography{wise}

\label{lastpage}
\end{document}